# Income Inequality and the Oil Resource Curse †

Osiris J. PARCERO [a]

Elissaios PAPYRAKIS [b, c]

[a] *Department of Economics and Finance, United Arab Emirates University, United Arab Emirates*

[b] *Institute for Social Studies (ISS), Erasmus University Rotterdam, Kortenaerkade 12, 2518 AX The Hague, The Netherlands*

[c] *School of International Development, University of East Anglia, Norwich, UK*

**Abstract**

Surprisingly, there has been little research conducted about the cross-country relationship between oil dependence/abundance and income inequality. At the same time, there is some tentative evidence suggesting that oil rich nations tend to under-report data on income inequality, which can potentially influence the estimated empirical relationships between oil richness and income inequality. In this paper we contribute to the literature in a twofold manner. First, we explore in depth the empirical relationship between oil and income inequality by making use of the Standardized World Income Inequality Database – the most comprehensive dataset on income inequality providing comparable data for the broadest set of country-year observations. Second, this is the first study to our knowledge that adopts an empirical framework to examine whether oil rich nations tend to under-report data on income inequality and the possible implications thereof. We make use of Heckman selection models to validate the tendency of oil rich countries to under-report and correct for the bias that might arise as a result of this – we find that oil decreases inequality with the exception of the very oil-rich economies.

JEL *classification*: D63, N50, O57

**Acknowledgements:** We would like to thank Abdul Rashid Faizi, Shahabuddin Abdul Rouf and Enayatullah Abdal Ghafoor for his excellent research assistance, as well as the editor and three anonymous referees for their very helpful comments on the paper. We are also grateful to Anke Hoeffler and Michael Ross for comments on an earlier draft. All remaining errors are ours.

† Correspondence: Osiris Jorge Parcero, Department of Economics and Finance, College of Business and Economics, United Arab Emirates University, osirisjorge.parcero@gmail.com.

**Highlights:**

- **Oil resources decrease inequality for moderate levels of oil affluence**
- **Oil-rich countries under-report data on income inequality**
- **Our results are robust once correcting for the sample selection bias**

# Income Inequality and the Oil Resource Curse

**Abstract**

Surprisingly, there has been little research conducted about the cross-country relationship between oil dependence/abundance and income inequality. At the same time, there is some tentative evidence suggesting that oil rich nations tend to under-report data on income inequality, which can potentially influence the estimated empirical relationships between oil richness and income inequality. In this paper we contribute to the literature in a twofold manner. First, we explore in depth the empirical relationship between oil and income inequality by making use of the Standardized World Income Inequality Database – the most comprehensive dataset on income inequality providing comparable data for the broadest set of country-year observations. Second, this is the first study to our knowledge that adopts an empirical framework to examine whether oil rich nations tend to under-report data on income inequality and the possible implications thereof. We make use of Heckman selection models to validate the tendency of oil rich countries to under-report and correct for the bias that might arise as a result of this – we find that oil is associated with lower income inequality with the exception of the very oil-rich economies.

1. INTRODUCTION

In recent years there has been a fast expanding literature researching the links between resource abundance and several measures of economic performance. Much of the so-called *resource curse* literature has developed theoretical and empirical research explaining the negative correlation observed between several measures of mineral abundance and long-term economic growth (Andersen and Aslaksen, 2008; Arezki and van der Ploeg, 2010; Baggio and Papyrakis, 2010; Caselli and Cunningham, 2009; Gylfason and Zoega, 2006; Kolstad, 2009; Murshed and Serino, 2011; Papyrakis and Gerlagh, 2004, 2007; Papyrakis, 2011, 2014; Sachs and Warner 1995, 1997, 1999a, 1999b, 2001). Much of this literature (to which this paper belongs) pays particular attention to oil and its correlates (e.g. for the case of conflict, see Lujala, 2010, gender inequality, see Ross, 2008, bureaucratic efficiency, see Goldberg et al., 2008).

Several explanations of the underperformance of oil rich economies have been provided in the literature. A first stream of the literature has focused on political economy explanations associating oil with the presence of inferior institutions and rent-seeking competition (Bjorvatn and Naghavi, 2011; Bjorvatn and Selvik, 2008; Bulte et al., 2005; Dalmazzo and De Blasio, 2003; Papyrakis et al., 2016; Torvik, 2002; Wick and Bulte, 2006). Competition for natural resource rents might also link to violent conflict, particularly in the case of ethnically fragmented societies (see Brunnschweiler and Bulte, 2009; Dixon, 2009; Olsson, 2007). A second branch of the literature looks at *Dutch Disease* explanations of poor economic performance (Beine et al., 2010; Cherif, 2013; Corden, 1984; Corden and Neary, 1982; Pegg, 2010; Papyrakis and Raveh, 2014; Torvik, 2002). In this context, mineral exports can be associated with both a relocation of production factors from various sectors towards the mineral sector as a result of wage premia in the latter (i.e. the so-called resource movement effect), as well as inflationary pressures and loss of competitiveness in exporting industries (i.e. the so-called spending effect).

While the resource curse literature initially focused attention on economic growth, it gradually broadened its scope to other development variables. For example, Bulte et al. (2005) and Daniele (2011) demonstrated that mineral resource dependence is associated with lower values of the *Human Development Index* (a composite development index of life expectancy, education and GDP per capita), undernourishment, higher child mortality and limited access to safe water. Ross (2008) claimed that oil dependence is associated with gender inequality measured by reduced female political representation and labour participation. Mineral-rich countries are also expected to be characterised by lower genuine savings (i.e. net total investment in physical, natural and human capital), that is often used as a measure of long-term (weak) sustainability (assuming that different forms of capital are perfectly substitutable, see Atkinson and Hamilton, 2003; Boos and Holm-Müller, 2012; Dietz and Neumayer, 2007). There is also some tentative evidence of a poor empirical track record of poverty alleviation in mineral dependent economies (see Pegg, 2006).

Surprisingly, though, there has been little research conducted about the relationship between oil dependence/abundance and income inequality. Oil rents can, in principle, link to lower income inequality if they encourage redistribution that favours low-income groups. On the other hand, they might relate to greater income inequality if they become concentrated in the hands of political elites or geographical regions (we expand on this further in Section 2). In this paper we contribute to this strand of the literature in a twofold manner. First, we explore in depth the empirical relationship between oil abundance/dependence and income inequality by making use of the Standardized World Income Inequality Database (SWIID) developed by Solt (2009). This is the most comprehensive dataset on income inequality providing comparable data for the broadest set of country-year observations. While our primary focus is to shed light on the links between oil and income inequality, our empirical specifications also control for other variables that have been found to influence income inequality in the literature. Second, this is the first study to our knowledge that adopts an empirical framework to examine whether oil rich nations tend to under-report data on income inequality and the possible implications thereof. We make use of Heckman selection models to validate the tendency of oil rich countries to under-report and correct for the bias that might arise as a result of this – we find that oil is associated with lower income inequality with the exception of the very oil-dependent economies.

The next section is devoted to the theoretical mechanisms that link income inequality to the presence of oil, as well as other possible explanatory factors. In the same section, we also discuss how oil dependence/abundance and other factors can influence the reporting behaviour of countries (regarding data on income inequality). The theoretical section will provide the justification behind the empirical specifications that are tested in subsequent sections. Section 3 presents our empirical analysis on income inequality and oil abundance. Section 4 focuses on the under-reporting behaviour of oil rich nations and presents a series of Heckman selection models that allow to correct for the bias that might arise from such under-reporting. Section 5 concludes.

## 2. INEQUALITY AND OIL: THE THEORY

In this section we discuss the theoretical mechanisms that are likely to link income inequality to the presence of oil, as well as other possible explanatory factors. We also comment on how these variables may not only relate to the level of income inequality, but also to the reporting of income inequality. The theoretical mechanisms presented will then shape the specifications that will be empirically tested in Sections 3 and 4 of the paper.

*Oil*

Oil rents may, in principle, be associated with lower income inequality if the revenues become redistributed equitably and possibly target lower income groups. On the other hand, an expansive oil sector may relate to greater income inequality by reducing production in the non-oil economy via Dutch

Disease effects, by inducing rent-seeking behaviour and disproportionately benefiting specific interest groups (particularly in ethnically fragmented societies, see Fum and Hodler, 2010) and/or disadvantaging the oil-scarce regions within the country (Ross, 2007). Gylfason and Zoega (2003) mention that resource abundance may correlate positively with income inequality when the distribution of natural capital is more unequal compared to other forms of capital in the economy. In an earlier paper, Leamer et al. (1999) suggest that the availability of natural resources (primarily land) relates to lower human capital accumulation, a diversion of physical capital away from manufacturing and higher levels of income inequality.

Ross (2007) claims that the Gini coefficient (the typical measure of income inequality) tends to be uncorrelated with mineral dependence, although he acknowledges that this might be driven by a sample bias. He claims that mineral rich nations tend to under-report data on income inequality, which can potentially influence the estimated empirical relationships between mineral dependence and income inequality (Ross, 2007). Williams (2011) suggests that governments in oil rich countries generally lack transparency and are reluctant to reveal sensitive information related to income inequality. Several other papers in the literature also suggest that oil-rich countries suffer from limited transparency (e.g. in terms of disclosing fiscal information; see de Renzio, 2009; O'Lear, 2011; Kalyuzhnova, 2011) – a phenomenon which could possibly also extend to the provision of inequality-related information. In oil rich nations there is generally lower dependence on taxes and heavier reliance on resource rents, which possibly reduces citizen demand for government accountability and transparency (e.g. see Ross, 2001; Sandbu, 2006).

*Income per Capita*

Some scholars hypothesise that richer countries are characterised by more equal income distributions, given their increased capacity for redistribution and relative prominence of the more labour-intensive service sector (Choi, 2006; Ravallion, 2010). The demand for egalitarian policies that place constraints on government behaviour and redistribute wealth away from political elites can also increase at higher income levels (Robinson and Acemoglu, 2002). Alternatively, poorer economies may suffer less from income inequality given the relative even distribution of income in predominantly agrarian societies. A non-linear relationship (where inequality first increases as income rises, but then falls for subsequently high levels of income; often referred to as the Kuznets curve) can also be possible as a result of the structural transformation of economies at different stages of development (with possibly higher levels of inequality at intermediate levels of GDP per capita, when the economy begins to industrialise; see Barro, 2000).

Income levels may also be associated with the extent of reporting inequality data. Other things equal, richer economies are likely to have better equipped administrations and statistical agencies to frequently collect and report data (on inequality, as well as other variables; see Williams, 2011). Furthermore, there are many empirical studies that demonstrate that the level of economic development is positively associated with several institutional/governance variables that are likely to matter for the release of information – e.g. richer nations tend to be more accountable to their electorates and are characterised by a more transparent public bureaucracy (see Goel and Ram, 2013; Paldam, 2002).

*Institutions*

Good institutions that support government accountability are likely to correlate with lower income inequality. Lee (2005) claims that fully institutionalised democracies are characterised by lower income inequality as a result of successful targeted redistribution – "democratic political mechanisms enable state institutions to be more responsive to the demands of the lower classes and more committed to achieving better distributional outcomes". In more authoritarian regimes, governments may use the public sector and fiscal mechanisms to selectively support industries and lobby groups with vested interests in government policies. Democratic institutions support the establishment of trade unions and political parties that represent the lower and middle classes and the expansion of suffrage has been historically associated with decreases in income inequality (see Reuveny and Li, 2003).

Strong democratic institutions may also correlate positively with the tendency to report data on income inequality. Williams (2009; 2011) claims, for instance, that good institutions that place constraints on the executive branch of the government are associated with higher levels of transparency and a higher flow of information released to the public by the government.

*Agriculture*

One might expect a negative correlation between agriculture and income inequality to the extent that income is more equally distributed across agrarian economies (e.g. see Andres and Ramlogan-Dobson, 2011; Chong, 2004). Particularly in the developing world, agriculture tends to be rather labour-intensive, with agricultural income distributed more equally across the population compared to income accruing from other sectors.

Could agriculture also play a role in terms of explaining variation in reporting behaviour across countries? Williams (2009) finds that countries that rely on agriculture tend to release more information to the public (while the opposite holds for the countries that rely on minerals). Bulte et al. (2005) also find a positive statistical relationship between agriculture and government effectiveness (with the latter capturing the overall quality of the civil service, which is likely to be closely associated with an adequate provision of information by the public administration). For this reason, we include agriculture as a control variable when we attempt to explain the variation in reporting behaviour across countries.

*Ethnic Fractionalisation*

Ethnic heterogeneity might relate to greater income inequality as a result of competitive rent-seeking across ethnic groups and selective redistributive government policies (Easterly and Levine, 1997; Milanovic, 2003). In general, people might be more averse to redistributive policies in places of higher ethnic diversity (Clarke et al., 2006).

The correlation between ethnic fractionalisation and the extent of data reporting (on inequality) is of an ambiguous sign. Fractionalised economies may be less transparent, particularly in the case of disclosing sensitive information that are likely to reveal the extent of (ethnically-based) income inequality (see Mauro, 1995). On the other hand, there is some contradicting evidence suggesting that governments in mineral-rich countries may actually make a concerted effort to improve transparency in the presence of an ethnically fractionalised population, in order to tackle mineral-induced rent-seeking (and for example, as a result of this, they are more likely to participate in the Extractive Industry Transparency Initiative that aims at improving transparency in the extractive sector, see Pitlik et al., 2010).

*Trade Openness*

The relationship between trade openness and income inequality is also ambiguous. Several theoretical papers suggest that trade openness and globalisation may either associate positively (Monfort and Nicolini, 2000; Paluzie, 2001) or negatively (Alonso-Villar, 2001; Behrens et al., 2007) with (regional) income inequality depending on the spatial implications of trade integration (i.e. patterns of internal dispersion or agglomeration of economic activity and comparative advantage). Empirical evidence is also mixed with some studies pointing to a positive (Ezcurra and Rodríquez-Pose, 2013; Rodríquez-Pose, 2012) and some to a negative (Asteriou et al., 2014; Zhou et al., 2011) link between trade expansion and income inequality (with the relationship largely depending on the sample selection and empirical specification).

According to Williams (2011), trade openness may also link to reporting behaviour at the country level. Countries more open to trade may have a tendency to release more information a. because their firms and individuals are exposed to norms of information disclosure of trade partners and might demand a similar level of transparency and b. because this can act as a positive signal to foreign investors. The hypothesis of whether this might also extend to the case of inequality-specific information will be tested in the empirical chapter that follows (for example, some countries that are open to trade may have a disincentive to disclose information on income inequality to their trade partners, if high income inequality is likely to act as a deterrent to foreign investment, e.g. because of income inequality being typically associated with higher incidences of crime and political instability).

## 3. INEQUALITY AND OIL: THE EMPIRICS

In this section we explore the association of income inequality with oil, as well as with a vector of other explanatory variables that have been found to be important in the literature. We rely on cross-country panel regressions to draw empirical estimations for these underlying relationships. Table 1 lists all variable descriptions, data sources and corresponding descriptive statistics. A matrix reporting pairwise correlations between all dependent and explanatory variables in our analysis is presented in Appendix 1 (for all countries, irrespective of reporting behaviour) and Appendix 2 (only for the reporting countries). Our initial empirical specification is of the following form:

$$Gini_{it} = \alpha_1 Oil_{i(t-5)} + \boldsymbol{\alpha}'_2 \boldsymbol{Z}_{it} + \boldsymbol{\alpha}'_3 \boldsymbol{R}_i + \boldsymbol{\alpha}'_4 \boldsymbol{T}_t + \varepsilon_{it}, \qquad (1)$$

where $Gini_{it}$ is the Gini coefficient of (net) income inequality for country $i$ at time $t$, $Oil_{i(t-5)}$ refers to our measure of oil richness (5-year lagged values), $\boldsymbol{Z}_{it}$ is a vector of control variables found to correlate with income inequality in the literature, regional dummies and time effects are captured by the vectors $\boldsymbol{R}_i$ and $\boldsymbol{T}_t$ respectively, and $\varepsilon_{it}$ corresponds to the error term[1]. One needs to keep in mind that an omitted variable bias can either over or under-estimate the coefficient of oil if the latter is correlated with unobserved characteristics that have not been accounted for.

Insert *Table 1*

We estimate equation (1) using pooled OLS regressions with country observations from an unbalanced panel (for the years 1975-2008). We opted for pooled OLS estimations, given that it is customary in the empirical literature to treat panels as extended cross-sectional datasets when pursuing baseline comparisons against panel Heckman selection models (which in effect also make use of pooled OLS procedures; e.g. see Dastidar, 2009; Dutt and Traca, 2009; Tang and Wei, 2009; Wooldridge, 1995; Zhou et al., 2011). For several specifications we also present the corresponding fixed effects estimates, although these tend to be less efficient for variables with little variation over time (which is the case for several of our explanatory variables, such as the measures of oil abundance/dependence, institutions, and GDP per capita levels, which fluctuate little from one year to the next; see Halaby, 2004, Hsiao, 2007 and Neumayer, 2004 for an elaborate discussion). Fixed-effect estimations tend to overinflate the standard errors of the coefficients corresponding to variables with little time variation. In any case, both estimation techniques produce similar qualitative results with respect to the correlation between oil and income inequality.

We present our first empirical estimations in Table 2 (using pooled OLS). Our dependent variable is the Gini coefficient of net income inequality (that is, public redistribution in the form of taxes and fiscal transfers is taken into account when index values are calculated). The coefficient ranges between 0 and 100, with larger values corresponding to more unequal income distributions. Data on the Gini coefficient come from the Standardized World Income Inequality Database (SWIID) developed by Solt (2009), which provides a very wide coverage of comparable income inequality data across countries (for 173 countries between 1960 and 2009). The SWIID dataset standardises data coming from multiple sources (e.g. the United Nations University's World Income Inequality Database, the OECD Income Distribution Database and the Socio-Economic Database for Latin America and the Caribbean by CEDLAS and the World Bank, as well as data from several national statistical offices). Observed values of Gini lie between 15 and 75. For all specifications, we include the 5-year lagged level of *Income* per capita (in logs), to control for any potential link between the level of economic development and income inequality (observed values lie between 5.7 and 11.0). Data on GDP per capita are provided by the World Development Indicators (World Bank, 2014). Our data provide empirical support (significant at the 1% level) for the first hypothesis pointing to a negative correlation between GDP per capita levels and the Gini coefficient. We also experimented with the quadratic form of income per capita to check for the validity of the Kuznets curve, that assumes an inverse U relationship between income inequality and the level of economic development. There was only weak statistical support for the quadratic form, which is in line with many other studies who also find that multicountry data do not support the Kuznet's hypothesis (Anand and Kanbur, 1993; Deininger and Squire, 1998; Andres and Ramlogan-Dobson, 2011).

---

[1] Appendix 3 provides the number of observations per country (based on our key specification (1) of Table 3) as well as the corresponding country-specific mean values for the different oil wealth variables.

In Column (1) of Table 2 we include the (5-year lagged) level of oil dependence (*Oil rents*) as an additional explanatory variable. We measure oil dependence as the value of annual oil rents in GDP (data are provided by the World Development Indicators of the World Bank (2014)) – the observed values of *Oil rents* lie between 0 and 82.07%. The coefficient of oil rents is a focal point of our analysis. Column (1) points to a negative and statistically-significant (at the 5% level) correlation between the (net) Gini coefficient and oil (i.e., income inequality is lower in oil-rich economies).

Insert *Table 2*

In Column (2) we test for a non-linear relationship between oil and inequality. Some papers in the resource curse literature suggest that a negative relationship between oil and other development outcomes holds only for sufficiently high levels of oil dependence (e.g. see Mehrara, 2009, for the case of growth and Crivelli and Gupta, 2014, for the case of tax collection). The linear term of our oil variable suggests that oil is associated with lower income inequality (e.g. if oil production accounts for an extra 10% in GDP, this corresponds to a lower Gini coefficient by approximately 1.2 units, assuming that countries start from very low levels of oil dependence) – the quadratic term is positive (although statistically insignificant), suggesting a reversal of sign; oil relates to greater inequality for very high levels of oil dependence (as a matter of fact, when oil rents account for more than 30% of GDP, as in the case of Venezuela for several years).

Column (3) expands the set of explanatory variables by including the Polity 2 variable from the Polity IV Project (variable *Institutions*) as a measure of institutional quality and democratic accountability of the political system (Marshall and Jaggers, 2009). The variable is arguably the most commonly used source of information for capturing cross-country variation in democratic (vs. authoritarian) governance (Plümper and Neumayer, 2010). Higher values of the index correspond to more extensive democratic governance (the index takes values between -10 and 10). We find that better institutions correspond to lower income inequality – for instance a positive difference of 19 units in the index (e.g. the difference between the most authoritarian country in the sample, Belarus, and the most democratic one, Sweden) relates to a lower Gini coefficient by approximately 5.7 units (in the 0-100 scale). Both the linear and quadratic terms of the oil variable appear to be statistically significant (at the 10% and 5% level respectively), suggesting that oil is associated with lower inequality (with the exception of highly oil dependent economies, where oil rents account for more than 52% of total income).

In Column (4) we enrich our empirical specification by including an additional regressor that has been found to correlate with income inequality in the literature, i.e. the share of agriculture in GDP (*Agriculture*). Data on the value of agricultural production are from the World Development Indicators (World Bank, 2014) – observed values lie between 0.06 and 73.4%. We find that the GDP share of agriculture is negatively correlated with income inequality as predicted by theory – when agriculture accounts for an extra 10% in GDP, this corresponds to a drop in the Gini coefficient by approximately two units.

In Column (5) of Table 1 we also introduce an index of ethnic fractionalisation. We make use of the ethnic *fractionalisation* index by Montalvo and Reynal-Querol (2005), which captures the probability of two randomly chosen individuals from the general population belonging to different ethnic groups. The index is of the following form: $fractionalisation = 1 - \sum_{i=1}^{N} \pi_i^2$, where $\pi_i$ stands for the proportion of the total population belonging to the *i-th* ethnic group and *N* stands for the number of groups. The fractionalisation index approaches unity as the number of different ethnic groups in the economy increases (and takes the value of zero for a perfectly homogenous society) – observed values lie between 0.01 and 1. The index by Montalvo and Reynal-Querol (2005) is one of the most commonly used proxies of ethnic fragmentation in the economic literature (e.g. see Akdede, 2010; Cole et al., 2013; Papyrakis, 2013) and largely correlates with other indices that use slightly different levels of disaggregation amongst ethnic groups (e.g. the ones by Alesina et al., 2003). As it is common in cross-country empirical analysis, the index of fractionalisation enters the regressions as a time invariant variable (it is customary in the economic literature to treat ethnic diversity as a non-time-varying variable due to limited data availability; e.g. see Arezki and Brückner, 2012; Montalvo and Reynal-Querol, 2005). We find ethnic heterogeneity to be positively associated with income inequality (and statistically significant at the 1% level) – a relatively ethnically fractionalised country (e.g. South Africa, with a score of 0.90) in comparison to a relatively ethnically homogeneous nation (e.g. Finland, with a score of 0.01) is expected to have a Gini coefficient that is larger by approximately 8.06 units.

Column (6) introduces *Openness* as an additional explanatory factor behind cross-country variation in income inequality (measured by the 5-year lagged values of exports and imports in GDP; data are provided by the World Development Indicators (World Bank, 2014)) – observed values lie between 6 and 444%. We find trade openness to have a negative correlation with income inequality (although the corresponding coefficient is statistically insignificant). For this richer specification, we again find that oil dependence is associated with lower income inequality, unless the country is highly oil dependent (with a mineral share in GDP above 25%). This fuller specification will become our main specification for the rest of the analysis.

Column (2) of Table 3 replicates the richer specification (6) of Table 2 (which appears also in Column (1) of Table 3 for the convenience of comparison) using a fixed effects estimation[2]. Results are qualitatively similar although some variables (e.g. institutions, agriculture, income) become less statistically significant (in line with our earlier concerns regarding fixed-effects estimations of specifications that include variables with little time variation; please note that ethnic fractionalisation drops out being time-invariant). The coefficients for the oil terms remain statistically significant at 1%, with the threshold level of oil dependence above which oil resources correspond to greater income inequality being equal to 28% (share of oil output in GDP).

Since the seminal work by Brunnschweiler and Bulte (2008), it is customary to distinguish between 'resource dependence' and 'resource abundance' indices, with the former measuring the value of resource rents as a share of economic activity (e.g. GDP, exports, etc.) and the latter in terms of population (i.e. a rather exogenous variable, less likely to be influenced by natural resources, should appear in the denominator). Several studies have found that any resource curse evidence disappears when one uses indices of mineral wealth in per capita terms rather than as a share of overall economic activity (e.g. see Brunnschweiler and Bulte, 2008; Cavalcanti et al., 2011; Stijns, 2006). In our analysis so far we used the GDP share of oil rents to measure oil dependence. Previous studies criticised this measure for being endogenous and associated with unobserved development characteristics (Brunnschweiler and Bulte, 2008; van der Ploeg 2011); in our case, this implies that our measure of oil richness can potentially be endogenous (as a result of being influenced by economic variables) and, as a result, our estimates may suffer from an endogeneity bias. For this reason we also check for the robustness of our main specification by experimenting with alternative measures of *oil abundance*: i.e. a. the 5-year lagged value of annual oil rents in per capita terms (in thousands of constant 2005 US dollars, rather than as a share of GDP; data are available from the World Bank, 2014 and observed values lie between 0 and 35.86 thousand US dollars) and b. the 5-year lagged value (in tens of thousands of constant 2005 US dollars) of the per capita value of known oil reserves (which being a stock measure, as opposed to the previously used flow measures of annual oil rents, is less vulnerable to endogeneity concerns – data are available from the World Bank, 2015 and observed values lie between 0 and 21.69). The data on the stock values of oil assets are, though, limited (available for 3 years; namely 1995, 2000 and 2005). Columns (3) and (4) of Table 3 replicate Columns (1) and (2) of Table 3 for the case of oil rents in per capita terms (pooled OLS and fixed effects respectively). Also in the case of our new 'oil abundance' measure (variable *Oil rents pc*), oil is associated with lower income inequality, unless the country is extremely oil abundant (with a value of oil abundance approximately five to seven standard deviations above the mean respectively; see Columns (3) and (4)). Columns (5) and (6) of Table 3 do the same for our second 'oil abundance' proxy (the value of known oil reserves in per capita terms; variable: *Oil reserves pc*). Results are very similar. Oil reserves are associated with lower income inequality with the exception of those very oil abundant economies (e.g. Venezuela).

---

[2] Table A1 in Appendix 4 re-estimates all columns of Table 2 using fixed effects.

Insert *Table 3*

Table 4 replicates Columns (1) and (2) of Table 3 by substituting the original institutional variable with three institutional dummies: i.e., *Institutions (very bad), Institutions (bad), Institutions (average)*, that take a value of 1 when our original institutional variable (*Institutions*) takes values between -10≤*Institutions*≤-5, -5<*Institutions*≤0 and 0<*Institutions*≤5 respectively (in other words, *Institutions>5,* i.e. the category with the highest scores for institutions*,* becomes the omitted category as it represents the dominant, most populous, group in our sample; for a discussion see Allen, 1997, p.138). These four ranges correspond to the four categories proposed by the Polity IV project, where the lowest scores in the range [-10, -5] characterise *autocracies* (i.e. a system of governance where the power is concentrated primarily in the hands of very few people, as in the case of absolute monarchy or a dictatorship), scores in the range (-5, 0] characterise *closed anocracies* (i.e. regimes that incorporate both autocratic and democratic elements, although any political competition is limited to dominant elite groups), scores in the range (0, 5] characterise *open anocracies* (i.e. regimes similar to closed anocracies, with the key difference being that political competition extends to a broader range of groups) and scores above 5 characterise *democracies* (i.e. systems of governance where the supreme power is vested in the people, through free and fair elections). Results (regarding the association between oil and inequality) are very similar, suggesting that oil dependence relates to lower income inequality (with the exception of the very oil-rich economies). The results (Column (1) of Table 4) suggest that, in comparison to the reference group of democratic countries, it is the countries in the middle range of democratic institutions (i.e. bad and average institutions) that are characterised by higher income inequality – autocracies (i.e. the countries with very bad institutions) do not appear to suffer from higher inequality (this is in line with Chong, 2004; Bourguignon and Verdier, 2000, who claim that at intermediate stages of democratisation it is mainly the middle classes that benefit rather than the poor, corresponding hence to an increase in income inequality).[3] Column (2) replicates the specification using fixed effects – the institutional dummies are not statistically significant (with the exception of the average institutions dummy, which is only significant at the 10% level), possibly as a result of their very limited time variation within the sample (results are also very similar when replicating the regressions for our alternative measures of oil wealth).

Insert *Table 4*

As an additional robustness check we replicate our richest empirical specification (for different proxies of oil wealth) by substituting *income* (GDP per capita) with *latitude* (data on latitude by Hall and Jones, 1999, observed values between 1 and 64 degrees); the pooled OLS results are presented in the first three columns of Table 5. Proximity to the tropics (captured by latitude) has been extensively used in empirical cross-country analysis as a proxy that can address the possible endogeneity of the levels of economic development (with proximity to the tropics determining income levels via the adverse health environment, e.g. see Angeles and Neanidis, 2015 and Pellegrini, 2011). Latitude is strongly and negatively correlated with income inequality; i.e. poorer nations closer to the tropics suffer from greater income inequality. There is little change in the estimated coefficients for all oil wealth proxies – oil richness is associated with lower income inequality with the exception of the very oil abundant/dependent nations. Another exogenous geographical variable that correlates strongly with the level of economic development is landlockedness (e.g. see Henderson et al., 2001; Sachs and Warner, 1997) – Columns (4)-(6) of Table 5 replicate the same specifications with landlockedness in place of income per capita (data on landlockedness by CIA, 2014). Landlocked nations tend to be characterised by lower levels of income inequality, although the effect is not statistically significant. Once again, oil abundance/dependence correlates negatively with income inequality with the exception of the very oil-rich economies.

---

[3] We also experimented with the quadratic form of institutions for all regressions, but found very weak support of a quadratic, inverse-U relationship; the linear specification, instead, always points to a statistically-significant negative relationship, suggesting that while inequality peaks at intermediate stages of democratization, democracies generally have the lowest inequality scores on the whole. In any case, irrespective of the use of quadratic or linear terms, the coefficients of the oil terms remains very similar in size and statistical significance.

Insert *Table 5*

### 4. OIL, UNDER-REPORTING AND HECKMAN CORRECTION

The statistical analysis presented in Section 3 can potentially provide erroneous conclusions to the extent that the statistical sample is non-random. In other words, it might be the case that the fact that some countries might provide no data on inequality for particular years (and hence are not included in the sample) is not unsystematic and can instead relate to some underlying factors (that we touched upon in Section 2). The two-step method of the Heckman correction model provides the means to correct for such non-randomly selected samples. The Heckman correction model allows us to use the limited non-random sample (i.e. the censored data, using the Heckman jargon) to draw inferences also for those countries for which we have missing values (the non-reporting ones) – what we need in order to do this, is to find out whether there is a statistical pattern explaining why a country might report or not data on inequality at a given time.

In an earlier study, Ross (2007) claimed that mineral rich countries tend to under-report data on income inequality (and are hence likely to be under-represented in the sample). Such under-reporting and corresponding sample bias can influence the estimated empirical relationship between measures of oil wealth and income inequality to the extent that under-reporting is not random and correlates with oil abundance/dependence and other underlying factors. In this section we estimate a series of Heckman selection models that allow us to correct for the bias that might arise from such under-reporting when estimating the relationship between inequality and oil. The first stage of the Heckman selection model estimates a *selection equation*, where the propensity to report (or not) depends on a number of factors. In this step (which is equivalent to a Probit regression) we hypothesise that reporting on income inequality might be associated with the extent of oil abundance/dependence, amongst other explanatory variables. In effect, the dependent variable in the first stage of the Heckman model is a dummy variable that takes a value of 1 if the country is reporting data on income inequality for that particular year, and 0 otherwise and we wish to see whether this dummy variable significantly correlates with a vector of regressors. Naturally, some countries might under-report data on income inequality as a result of a broader tendency to under-report data of any kind, for example as a consequence of weak government administration (i.e. the tendency to under-report data on income inequality may reflect inadequate government capacity, or simply bad practice, rather than any intentional effort to conceal information that could be considered to be sensitive). For this reason we construct an index (*Data reporting index*) that proxies the overall tendency (capacity) of countries to report data of any kind[4] – we measure this as the number of data entries reported for any variables appearing in the World Development Indicators database for country *i* and year *t* (World Bank, 2014), expressed as a share of the number of (annual) data entries for the 'average reporting country' during the period of analysis. Observed values lie between 0.31 and 2.72. We expect that the *Data reporting index* is likely to correlate positively with the probability of disclosing data on income inequality.

---

[4] In addition this variable acts as an exclusion restriction. In the Heckman selection model an exclusion restriction is a variable that appears in the selection equation, but not in the outcome equation. The inclusion of such a variable is the most common means of solving the over-identification problem, see Fu and Mare (2004).

More specifically, the Heckman model considers that observations are ordered into two regimes. In the present context these regimes are defined by whether or not the country reports data on inequality. The first stage defines a dichotomous variable indicating the regime into which the observation falls:

$$R_{it}^* = \tau_1 Data\ Reporting\ Index_{it} + \tau_2 Oil_{i(t-5)} + \tau_3' Z_{it} + \tau_4' R_i + \tau_5' T_t + \epsilon_{it} \qquad (2)$$

$R_{it} = 1$ if $R_{it}^* > 0$ (in which case income inequality is observed as in eq (1) and

$R_{it} = 0$ if $R_{it}^* \leq 0$ (in which case data on income inequality are missing), (3)

where $R_{it}^*$ is a latent variable indicating the utility of reporting data on income inequality, $R_{it}$ is an indicator for the reporting behaviour (the dummy we described earlier), $Z_{it}$ is the vector of control variables already defined above, regional dummies and time effects are captured in the vectors $R_i$ and $T_t$ respectively (for the rest of the analysis we will omit these two vectors to simplify notation and assume they are part of vector **Z**), and $\epsilon_{it}$ is the error term. Keeping the exact same control variables of the vector Z in both the main regression, as well as the selection equation, is common practice (e.g. see Baudassé and Bazillier, 2014; Fleck and Kilby, 2010) - and as a matter of fact, it is advisable, given that the exclusion of some of the original control variables can lead to inconsistent estimates; for a discussion see Wooldridge (2012, p.619). We use the two-step method estimation which is based on the following conditional expectation[5]:

---

[5] The two-step estimation relies on a univariate normality assumption for $\epsilon_{it}$ and $v_{it}$ and is expected to be relatively more robust that the Maximum Likelihood estimation, which relies on a bivariate normality assumption.

$$E(Gini_{it}|R^*_{it} > 0, Oil_{i(t-5)}, \mathbf{Z}_{it}) = \beta_1 Oil_{i(t-5)} + \boldsymbol{\beta}_2 \mathbf{Z}_{it} + \beta_\lambda \lambda_{it}(\phi(\cdot)/\Phi(\cdot)) + v_{it}, \quad (4)$$

where $\lambda_{it}(\phi(\cdot)/\Phi(\cdot))$ is the inverse Mill's ratio, defined by the ratio of the density function of the standard normal distribution, $\phi$, to its cumulative density function, $\Phi$. The third term in (4) can be estimated by $\lambda_{it}(\hat{\tau}_1 Data\_reporting\_index_{it} + \hat{\tau}_2 Oil_{i(t-5)} + \hat{\boldsymbol{\tau}}'_3 \mathbf{Z}_{it})$, where $\hat{\tau}_1$, $\hat{\tau}_2$ and $\hat{\boldsymbol{\tau}}'_3$ are obtained by applying a Probit regression to (2). The regression of $Gini_{it}$ on $Oil_{i(t-5)}$ and $\mathbf{Z}_{it}$ and the generated regressor, $\lambda_{it}(\hat{\tau}_1 Data\ Reporting\ Index_{it} + \hat{\tau}_2 Oil_{i(t-5)} + \hat{\boldsymbol{\tau}}'_3 \mathbf{Z}_{it})$ yields a semi-parametric estimate of $(\beta_1, \boldsymbol{\beta}_2, \beta_\lambda)$.

Let's first start with the oil dependence variable we initially used in our analysis in Section 3 (i.e. the share of oil rents in GDP). The results of the first stage of the Heckman selection model are presented at the bottom panel of Table 6 (Column (1)), where the first regressor is the *Data reporting index* and the rest of the regressors appear in the same order as the ones present in the empirical specifications of Table 3 (i.e. the panel regressions that do not correct for any systematic selection bias). Some interesting findings are revealed. First, in line with Ross (2007), we find that oil rich nations tend to under-report on income inequality (a relationship that is statistically significant at the 1% level in all the specifications). This holds even when we control for the *Data reporting index*, which captures the overall tendency (capacity) of countries to report data of any kind (the effect is positive and statistically significant). Democratic accountability (*Institutions*), agriculture, income per capita and fractionalisation are also positively and significantly correlated with reporting behaviour (while trade openness has the opposite sign). It is important to note that the parameter λ (bottom panel of Table 6) appears to be significant in all regressions, suggesting that the null hypothesis of a selection problem is not rejected. In more technical terms, the null hypothesis that the two parts of the model are independent is rejected.

Insert *Table 6*

The results of the second stage of the Heckman Selection Model (top panel of Table 6) make inferences based on both the countries reporting data on inequality, as well as the non-reporting ones for which data were inferred based on the first stage. However, the coefficients of the Heckman's second stage cannot be directly interpreted as the corresponding marginal effects, which need to be calculated separately. Some earlier papers in the literature show unawareness of this issue, though the formula for the calculation of the marginal effect in the Heckman Selection Model has been gaining popularity. The formula is explained in detail in Green (2000), as well as in Hoffmann and Kassouf (2005). Moreover, a general version of it is proposed in Frondel and Vance (2009), which allows for the calculation of the marginal effect even in the presence of interaction terms. The marginal effects essentially capture both the direct link of the independent variable to income inequality (second stage of Heckman) as well as the indirect effect on the probability that an observation is part of the sample (first stage of Heckman). Moreover, there are two types of marginal effects in the Heckman Selection Model, the conditional and the unconditional ones. The conditional marginal effect is the one for the observed (uncensored) sample and so the most appropriate to be compared against the coefficients from the pooled OLS regressions (Section 3). The conditional marginal effects corresponding to Column (1) of Table 6 (for mean values of variables) have been calculated by using formula A1 in Appendix 5 (see Frondel and Vance, 2009, for a detailed description) and are presented next to the Heckman coefficients. Clearly, the marginal effects do not need to have the same magnitude or even the same sign as the second stage coefficients; the latter case occurs, for example, for the variable

*Openness*.[6] *Oil rents* are negatively associated with net income inequality (and are statistically significant at the 1% level) – this is also in line with our earlier findings in Section 3, suggesting that the empirical relationship between oil and lower inequality (below a certain level of oil dependence) is robust to any selection bias arising from the tendency of oil-rich countries to under-report data on income inequality (as a matter of fact, the marginal effect suggested by the Heckman model is a bit larger compared to the estimated effect of oil rents (at their mean) of the corresponding pooled OLS regression; i.e. of Column (1) of Table 3). *Fractionalisation* also appears to be positively and significantly linked to net income inequality (while the opposite holds for *agriculture*).

Figure 1 presents the marginal effects (blue dotted line) of oil rents (as a share of GDP) on income inequality for varying levels of oil dependence that correspond to Column (1) of Table 6 (one can see that the marginal effect of oil dependence on inequality is negative for countries with a share of oil rents in GDP below 23.5%). The confidence intervals (red dotted lines) around the marginal effects line determine the statistical significance of the marginal effect of oil (at each given level of oil dependence); the marginal effect is statistically significant at the 5% level when both the upper and lower bounds of the confidence intervals are below the zero line (i.e. for a share of oil rents in GDP below 19%). One can see, that, for countries where *Oil rents* are close to their sample-mean value of 3.7, a marginal increase in the share of oil rents in GDP by 1% is associated with lower income inequality (i.e. the Gini coefficient) by 0.539 units. The link between oil and lower income inequality decreases in magnitude as the share of oil rents in GDP increases (e.g. one can see from Figure 1 that the corresponding marginal effect is close to -0.3, when the share of oil rents in GDP is close to 12%).

Insert *Figure 1*

---

[6] The reason for this is clear in formula A1 of Appendix 5. Note that one important determinant of the marginal effect sign is the sign of $\delta(u_1)$.

Column (2) of Table 6 replicates the Heckman selection model for our first alternative 'oil abundance' measure, i.e. the value of oil rents in per capita terms. The conditional marginal effects (for mean values of variables) are also presented next to the Heckman coefficients. Results are in line with the earlier findings of Section 3 – although oil abundant countries tend to under-report data on inequality, the selection bias does not alter our earlier findings pointing to a negative relationship between oil and lower inequality (for moderate levels of oil rents in per capita terms). Figure 2 presents the marginal effects of oil abundance (*oil rents pc*) on income inequality for varying levels of oil abundance corresponding to Column (2) of Table 6 (the marginal effect of per capita oil rents on inequality is negative for countries with a moderate level of oil abundance; i.e. below 3.2 thousand US dollars per head in constant 2005 prices). One can see, that, for countries where *Oil rents pc* (per capita oil rents) are close to their sample-mean value of 0.25 (i.e. 250 US$), an increase in per capita oil rents by 100 dollars is associated with lower income inequality (i.e. the Gini coefficient) by approximately 0.55 units. Column (3) of Table 6 replicates the Heckman specification using our second 'oil abundance' proxy (the value of known oil reserves in per capita terms; variable: *Oil reserves pc*) – results are also in line with our earlier findings.

Insert *Figure 2*

In Table 7 we replicate the Heckman selection specifications of Table 6 for our three oil measures (oil rents in GDP, oil rents per capita, oil reserves per capita), using *latitude* in place of the *income* variable. The coefficients (and corresponding marginal effects) of the oil measures change very little. Figures 3 and 4 present the marginal effects for oil rents in GDP and per capita oil rents respectively, for varying levels of corresponding resource dependence (abundance). The results are in line with our earlier findings (i.e. oil resources are associated with lower income inequality, for moderate values of oil affluence).[7]

Insert *Table 7*

Insert *Figure 3*

Insert *Figure 4*

### 3. CONCLUSION

There has been an increasing interest in recent years in the relationship between oil and broader socio-economic development. Oil rents can, in principle, link to lower income inequality by magnifying the ability of governments to redistribute public revenues and improve the relative position of the economically disadvantaged. Surprisingly, though, the relationship between oil wealth and income inequality has been largely under-researched. In this paper we empirically explore in depth the relationship between oil abundance/dependence and income inequality by making use of the Standardized World Income Inequality Database (SWIID) and we pay particular attention to the tendency of oil rich nations to under-report relevant data. We make use of Heckman selection models to validate the tendency of oil rich countries to under-report and correct for the bias that might arise as a result of this – we find that oil resources are associated with greater income inequality only for the very oil-dependent economies (and have the opposite effect for moderate levels of oil dependence/abundance).

These findings have significant policy implications. With limited information about the link between oil and income inequality (and given the general presumption from the resource curse literature pointing to a negative relationship between oil resources and most development outcomes), policy-makers are likely to expect that oil rents are associated with a less equitable distribution of income. The tendency of oil rich economies to under-report data on income inequality might also hint on an intentional effort to conceal such sensitive information. Here, we have shown that oil resources in most cases are associated with a more equitable income distribution (and with greater income inequality only for the more extreme cases of oil abundance/dependence). For this reason, governments in modestly oil rich nations, as well as the international community (donors, international organizations), do not need to design a different set of distributive policies but rather pay more attention on how to utilize the oil rents against resource curse 'ailments' that are already well-established in the literature (such as the lack of economic diversification or excessive investment in 'white elephant'-type public investment projects).

---

[7] We also replicated the results using the three institutional dummy variables of Table 4, as well as for landlockedness in place of latitude. Results are in line with our earlier findings and are available from the authors upon request.

The question of what makes some countries more successful than others in managing their oil revenues is certainly one of the most fascinating economists can ask. Our analysis is simply a first step in exploring the intriguing relationship between oil and income inequality. Future research could attempt to disentangle in more detail the mechanisms (or the 'transmission channels' as commonly referred to in the literature) through which oil resources can influence the income distribution – for example by looking at income inequality levels across regions, gender groups or economic sectors. Another direction for future research would be to complement large-sample econometric studies with more case histories of economic policy in oil-rich countries (e.g. by looking at how decision-making and redistributive policies are shaped in the context of oil dependent economies).

TABLE 1. *List of Variables Used in Regression Analysis and Descriptive Statistics.*

| Variable Name | Variable Description and Data Source | Mean | Standard Deviation | Minimum/Maximum |
|---|---|---|---|---|
| *Gini* | Gini coefficient of income inequality (net of taxes and transfers). Index ranging between 0 and 100, with larger values corresponding to more unequal income distributions. Source: Solt (2009). | 37 [37] | 11 [11] | 15/75 [15]/[75] |
| *Income* | 5-year lagged real GDP per capita (constant 2005 US dollars). Source: World Bank (2014). | 9,090 [14,201] | 9,579 [9,567] | 309/59,239 [918]/[42,490] |
| *Log Income* | Natural logarithm of *Income* variable. Source: World Bank (2014). | 9.1 [9.6] | 9.2 [9.2] | 5.7/11.0 [6.8]/[10.7] |
| *Latitude* | The absolute value of the latitude of a country. Source: Hall and Jones (1999). | 23 [31] | 16 [19] | 1/64 [1]/[64] |
| *Landlocked* | Dummy variable taking a value of 1 if country landlocked, 0 otherwise. Data from the CIA (2014) World Factbook. | 0.18 [0.09] | 0.39 [0.28] | 0/1 [0]/[1] |
| *Oil rents* | 5-year lagged value of the share of annual oil rents in GDP (Source: World Bank (2014). | 3.7 [2.3] | 8.35 [5.14] | 0/82.07 [0]/[43.75] |
| *Oil rents pc* | 5-year lagged value (in thousands of constant 2005 US dollars) of annual oil rents in per capita terms (Source: World Bank (2014). | 0.25 [0.13] | 1.4 [0.45] | 0/35.86 [0]/[6.34] |
| *Oil reserves pc* | 5-year lagged value (in tens of thousands of constant 2005 US dollars) of the per capita value of known oil reserves. The value of oil reserves is calculated as the present value of expected rents from oil extraction, discounted at 4 percent. For data and discussion of methodology behind calculation, see World Bank (2015). | 0.45 [0.19] | 1.89 [0.65] | 0/21.69 [0]/[5.72] |
| *Institutions* | Polity 2 index (in the range between -10 to 10) from the Polity IV Project measuring the democratic accountability of the political system. Higher values corresponding to greater democratic governance. Source: Marshall and Jaggers (2009). | 2.7 [6.9] | 7 [5.1] | -10/10 [-9]/[10] |
| *Agriculture* | Share of agricultural production in GDP. Source: World Bank (2014). | 14.8 [17.6] | 14 [15.2] | 0.0610/73.4 [0.86]/[72.6] |
| Fractionalisation | Ethnic fractionalisation index (0-1 continuous scale). Source: Montalvo and Reynal-Querol (2005). | 0.47 [0.38] | 0.28 [0.27] | 0.01/1 [0.01]/[0.9] |
| *Openness* | The share of the value of exports and imports in GDP (5-year lagged values). Source: World Bank (2014). | 71 [72] | 47 [55] | 6/444 [12]/[444] |
| *Reporting behaviour* | 0-1 index measuring reporting behaviour of income inequality data (Gini, redistribution). A 0 value corresponds to non-reporting. Values calculated by authors based on data by Solt (2009). | 0.43 [1] | 0.5 [0] | 0/1 [0]/[1] |
| *Data reporting index* | A proxy of the overall tendency (capacity) of countries to report data of any kind. The index is measured as the number of data entries reported for any variables appearing in the World Development Indicators database for country *i* and year *t* (World Bank, 2014) as a share of the number of (annual) data entries for the 'average reporting country' during the period of analysis. | 1.27 [1.32] | 0.19 [0.14] | 0.31/2.72 [0.8]/[1.65] |

Note: This table is based on the observations included in specification (6) of Table 2. Descriptive statistics for the joint sample of both Gini reporting and non-reporting countries are outside the squared brackets. Descriptive statistics for the sample of reporting countries are inside squared brackets.

36TABLE 2. *Effects of Oil Dependence on Net Income Inequality (Pooled OLS)*

| Dependent variable: | Gini (1) | Gini (2) | Gini (3) | Gini (4) | Gini (5) | Gini (6) |
| --- | --- | --- | --- | --- | --- | --- |
| *Log Income* | -4.454*** (0.281) | -4.477*** (0.278) | -3.372*** (0.392) | -4.104*** (0.377) | -4.004*** (0.478) | -3.999*** (0.473) |
| *Oil rents* | -0.072** (0.029) | -0.123** (0.049) | -0.209*** (0.048) | -0.255*** (0.043) | -0.495*** (0.068) | -0.496*** (0.068) |
| *Oil rents (sq)* | | 0.002 (0.001) | 0.002** (0.001) | 0.003*** (0.001) | 0.010*** (0.002) | 0.010*** (0.002) |
| *Institutions* | | | -0.301*** (0.052) | -0.168*** (0.043) | -0.214*** (0.045) | -0.215** (0.046) |
| *Agriculture* | | | | -0.201*** (0.014) | -0.164*** (0.015) | -0.164*** (0.016) |
| *Fractionalisation* | | | | | 9.053*** (0.743) | 9.013*** (0.811) |
| *Openness* | | | | | | -0.068 (0.299) |
| $R^2$ adjusted | 0.563 | 0.565 | 0.608 | 0.663 | 0.735 | 0.734 |
| $N$ | 1935 | 1935 | 1592 | 1572 | 1348 | 1348 |
| Countries | 81 | 81 | 75 | 75 | 55 | 55 |

Note: Robust standard errors of coefficients in parentheses. Superscripts *, **, *** correspond to a 10, 5 and 1% level of significance. Year and regional dummies (Sub-Saharan Africa, East Asia, South Asia, Latin America) included in all specifications. A detailed description of all variables is provided in Table 1.

TABLE 3. *Effects of Oil Dependence/Abundance on Income Inequality (pooled OLS/ Fixed Effects)*

| Dependent variable: | Gini (OLS) (1) | Gini (FE) (2) | Gini (OLS) (3) | Gini (FE) (4) | Gini (OLS) (5) | Gini (FE) (6) |
|---|---|---|---|---|---|---|
| *Log Income* | -3.999*** (0.473) | 5.092* (2.991) | -3.589*** (0.488) | 5.730 (3.638) | -4.927*** (0.708) | 10.788*** (3.529) |
| *Oil rents* | -0.496*** (0.068) | -0.896*** (0.316) | | | | |
| *Oil rents (sq)* | 0.010*** (0.002) | 0.016*** (0.006) | | | | |
| *Oil rents pc* | | | -5.257*** (0.552) | -2.065** (0.937) | | |
| *Oil rents pc (sq)* | | | 0.851*** (0.134) | 0.415** (0.159) | | |
| *Oil reserves pc* | | | | | -5.022*** (0.713) | 2.599 (3.576) |
| *Oil reserves pc (sq)* | | | | | 0.749*** (0.162) | -0.257 (0.401) |
| *Institutions* | -0.215** (0.046) | 0.057 (0.147) | -0.189*** (0.045) | 0.081 (0.164) | -0.231*** (0.088) | -0.026 (0.128) |
| *Agriculture* | -0.164*** (0.016) | -0.301** (0.139) | -0.183*** (0.016) | -0.270* (0.157) | -0.229*** (0.025) | -0.212* (0.126) |
| *Fractionalisation* | 9.013*** (0.811) | | 7.229*** (0.780) | | 7.352*** (1.350) | |
| *Openness* | -0.068 (0.299) | 0.923 (1.374) | -0.054 (0.301) | 1.149 (1.528) | -0.781* (0.419) | -0.809 (1.590) |
| $R^2$ adjusted | 0.734 | 0.260 | 0.731 | 0.203 | 0.736 | 0.187 |
| $N$ | 1348 | 1348 | 1348 | 1348 | 582 | 582 |
| Countries | 55 | 55 | 55 | 55 | 54 | 54 |

Note: Robust standard errors of coefficients in parentheses. Superscripts *, **, *** correspond to a 10, 5 and 1% level of significance. Year dummies included in all specifications; regional dummies (Sub-Saharan Africa, East Asia, South Asia, Latin America) included in pooled OLS specifications. A detailed description of all variables is provided in Table 1.

TABLE 4. *Effects of Oil Dependence on Net Income Inequality (Institutional Dummy Variables)*

| Dependent variable: | Gini (OLS) (1) | Gini (FE) (2) |
|---|---|---|
| *Log Income* | -4.934*** [0.504] | 4.681 [2.964] |
| *Oil rents* | -0.473*** [0.071] | -0.890*** [0.321] |
| *Oil rents (sq)* | 0.008*** [0.003] | 0.016*** [0.006] |
| *Institutions (very bad)* | 0.270 [0.900] | -1.026 [2.405] |
| *Institutions (bad)* | 5.475*** [0.858] | -1.129 [2.011] |
| *Institutions (average)* | 3.607*** [0.894] | -2.141* [1.069] |
| *Agriculture* | -0.157*** [0.015] | -0.298** [0.136] |
| *Fractionalisation* | 8.304*** [0.810] | |
| *Openness* | -0.631** [0.314] | 0.944 [1.335] |
| $R^2$ adjusted | 0.742 | 0.271 |
| $N$ | 1348 | 1348 |
| Countries | 55 | 55 |

Note: Robust standard errors of coefficients in parentheses. Superscripts *, **, *** correspond to a 10, 5 and 1% level of significance. Year dummies included in all specifications; regional dummies (Sub-Saharan Africa, East Asia, South Asia, Latin America) included in the pooled OLS specification (1). A detailed description of all variables is provided in Table 1.

TABLE 5. *Effects of Oil on Net Income Inequality (Latitude/Landlocked; Pooled OLS)*

| Dependent variable: | Gini (1) (Latitude) | Gini (2) (Latitude) | Gini (3) (Latitude) | Gini (4) (Landlocked) | Gini (5) (Landlocked) | Gini (6) (Landlocked) |
|---|---|---|---|---|---|---|
| *Latitude / Landlocked* | -0.179*** (0.015) | -0.150*** (0.015) | -0.204*** (0.027) | -0.480 (0.600) | -0.155 (0.587) | -0.986 (1.096) |
| *Oil rents* | -0.561*** (0.061) | | | -0.544*** (0.069) | | |
| *Oil rents (sq)* | 0.011*** (0.002) | | | 0.011*** (0.003) | | |
| *Oil rents pc* | | -5.371*** (0.619) | | | -6.841*** (0.647) | |
| *Oil rents pc (sq)* | | 0.860*** (0.169) | | | 1.108*** (0.181) | |
| *Oil reserves pc* | | | -7.636*** (0.924) | | | -7.196*** (0.850) |
| *Oil reserves pc (sq)* | | | 1.412*** (0.250) | | | 1.138*** (0.207) |
| *Institutions* | -0.379*** (0.046) | -0.343*** (0.046) | -0.468*** (0.077) | -0.543*** (0.043) | -0.469*** (0.044) | -0.682*** (0.074) |
| *Agriculture* | -0.130*** (0.017) | -0.150*** (0.017) | -0.160*** (0.029) | -0.147*** (0.017) | -0.171*** (0.017) | -0.196*** (0.028) |
| *Fractionalisation* | 7.071*** (0.710) | 5.392*** (0.711) | 6.080*** (1.253) | 10.248*** (0.941) | 8.320*** (0.901) | 9.120*** (1.671) |
| *Openness* | -1.210*** (0.294) | -1.013*** (0.294) | -2.114*** (0.466) | -0.272 (0.319) | -0.237 (0.313) | -0.974** (0.461) |
| $R^2$ adjusted | 0.739 | 0.731 | 0.732 | 0.705 | 0.708 | 0.694 |
| $N$ | 1348 | 1348 | 582 | 1,348 | 1,348 | 582 |
| Countries | 55 | 55 | 54 | 55 | 55 | 54 |

Note: Robust standard errors of coefficients in parentheses. Superscripts *, **, *** correspond to a 10, 5 and 1% level of significance. Year and regional dummies (Sub-Saharan Africa, East Asia, South Asia, Latin America) included in all specifications. A detailed description of all variables is provided in Table 1.

40TABLE 6. *Oil, Under-reporting on Net Inequality and Heckman Correction*

| Dependent variable: Gini (net) (2$^{nd}$ stage) | (1) Oil rents | | (2) Oil rents pc | | (3) Oil reserves pc | |
|---|---|---|---|---|---|---|
| | Heckman | Marginal Effects | Heckman | Marginal Effects | Heckman | Marginal Effects |
| *Log Income* | -4.315*** (0.516) | -3.361*** (0.503) | -3.966*** (0.542) | -2.795*** (0.533) | -6.727*** (0.744) | -1.453 (1.032) |
| *Oil (several measures)* | -0.550*** (0.078) | -0.539*** (0.071) | -4.901*** (0.780) | -5.259*** (0.761) | -4.253*** (1.101) | -5.335*** (1.104) |
| *Oil (sq) (several measures)* | 0.012*** (0.003) | | 0.863*** (0.183) | | 0.704*** (0.254) | |
| *Institutions* | -0.248*** (0.050) | -0.199*** (0.048) | -0.241*** (0.051) | -0.181*** (0.048) | -0.384*** (0.080) | -.2665*** (0091) |
| *Agriculture* | -0.162*** (0.016) | -0.125*** (0.024) | -0.175*** (0.016) | -0.136*** (0.023) | -0.269*** (0.026) | -0.082 (0. 050) |
| *Fractionalisation* | 8.784*** (0.845) | 9.581*** (0.912) | 6.804*** (0.804) | 7.523*** (0.839) | 4.778*** (1.356) | 9.230*** (1.827) |
| *Openness* | 0.073 (0.289) | -0.187 (0.321) | 0.013 (0.289) | -0.246 (0.318) | -0.534 (0.455) | -2.377*** (0. 769) |
| Dependent variable: Reporting behaviour (1$^{st}$ stage) | | | | | | |
| *Data reporting index* | 1.961*** (0.251) | | 1.956*** (0.251) | | 2.669*** (0.485) | |
| *Income* | 0.920*** (0.058) | | 0.992*** (0.060) | | 1.197*** (0.111) | |
| *Oil (several measures)* | -0.046*** (0.012) | | -0.500*** (0.077) | | -0.309*** (0.103) | |
| *Oil (sq) (several measures)* | 0.001 (0.000) | | 0.013*** (0.003) | | 0.005 (0.011) | |
| *Institutions* | 0.047*** (0.006) | | 0.052*** (0.006) | | 0.027** (0.011) | |
| *Agriculture* | 0.035*** (0.004) | | 0.034*** (0.003) | | 0.042*** (0.006) | |
| *Fractionalisation* | 0.768*** (0.174) | | 0.760*** (0.164) | | 1.010*** (0.267) | |
| *Openness* | -0.251*** (0.066) | | -0.219** (0.066) | | -0.418*** (0.119) | |
| $\beta\lambda$ | -1.207*** (0.962) | | -1.375*** (0.958) | | -4.934*** (1.202) | |
| *Uncensored Observ.* | 1348 | | 1348 | | 582 | |
| *Censored Observ.* | 1655 | | 1655 | | 556 | |
| *Censored Countries* | 55 | | 55 | | 54 | |

Note: Robust standard errors of coefficients in parentheses. Superscripts *, **, *** correspond to a 10, 5 and 1% level of significance. Year and regional dummies (Sub-Saharan Africa, East Asia, South Asia, Latin America) included in all specifications. A detailed description of all variables is provided in Table 1. Calculations of the marginal effects and standard errors (inside parentheses) done through the 'delta method' by using the Stata command nlcom (based on the sample mean values of variables for each specification).

41TABLE 7. *Oil, Under-reporting on Net Inequality and Heckman Correction (Latitude)*

| Dependent variable: Gini (net) (2nd stage) | (1) Oil rents | | (2) Oil rents pc | | (3) Oil reserves pc | |
|---|---|---|---|---|---|---|
| | Heckman | Marginal Effects | Heckman | Marginal Effects | Heckman | Marginal Effects |
| *Latitude* | -0.161*** (0.017) | -0.222*** (0.020) | -0.132*** (0.001) | -0.202*** (0.021) | -0.214*** (0.023) | -0.183*** (0.028) |
| *Oil (several measures)* | -0.613*** (0.724) | -0506*** (0.069) | -5.008*** (0.001) | -4.678*** (0.760) | -7.271*** (1.114) | -6.358*** (1.101) |
| *Oil (sq) (several measures)* | 0.012*** (0.174) | | 0.782*** (0.001) | | 1.329*** (0.270) | |
| *Institutions* | -0.248*** (0.057) | -0437*** (0048) | -0.206*** (0.001) | -.425*** (0.049) | -0.569*** (0.090) | -0.437*** (0.073) |
| *Agriculture* | -0.086*** (0.014) | -0.145*** (0.020) | -0.103*** (0.001) | -0.164*** (0.020) | -0.162*** (0.024) | -0.120*** (0.031) |
| *Fractionalisation* | 7.231*** (0.786) | 5.825*** (0.974) | 5.459*** (0.001) | 3.978*** (0.957) | 4.715*** (1.301) | 5.546*** (1.355) |
| *Openness* | -1.373*** (0.271) | -1.484*** (0.333) | -1.235*** (0.001) | -1.457*** (0.349) | -2.129*** (0.448) | -2.181*** (0.468) |
| Dependent variable: Reporting behaviour (1st stage) | | | | | | |
| *Data reporting index* | 2.101*** (0.211) | | 2.257*** (0.001) | | 3.560*** (0.432) | |
| *Latitude* | 0.026*** (0.003) | | 0.027*** (0.001) | | 0.020*** (0.005) | |
| *Oil (several measures)* | -0.023*** (0.081) | | -0.047 (0.647) | | 0.284 (0.250) | |
| *Oil (sq) (several measures)* | 0.001 (0.007) | | -0.002 (0.912) | | -0.060 (0.059) | |
| *Institutions* | 0.079*** (0.005) | | 0.085*** (0.001) | | 0.085*** (0.010) | |
| *Agriculture* | 0.025*** (0.003) | | 0.024*** (0.001) | | 0.027*** (0.005) | |
| *Fractionalisation* | 0.589*** (0.154) | | 0.573*** (0.001) | | 0.537** (0.253) | |
| *Openness* | 0.046 (0.056) | | 0.086 (0.152) | | -0.034 (0.099) | |
| $\beta\lambda$ | 2.690*** (0.97) | | 2.879*** (0.001) | | -1.637 (0.152) | |
| *Uncensored Observ.* | 1348 | | 1348 | | 582 | |
| *Censored Observ.* | 1655 | | 1655 | | 556 | |
| *Censored Countries* | 55 | | 55 | | 54 | |

Note: Robust standard errors of coefficients in parentheses. Superscripts *, **, *** correspond to a 10, 5 and 1% level of significance. Year and regional dummies (Sub-Saharan Africa, East Asia, South Asia, Latin America) included in all specifications. A detailed description of all variables is provided in Table 1. Calculations of the marginal effects and standard errors (inside parentheses) done through the 'delta method' by using the Stata command nlcom (based on the sample mean values of variables for each specification).

FIGURE 1. *Marginal Effects of Oil on Gini (Net) for Different Values of Oil Dependence (Specification 1, Table 6)*

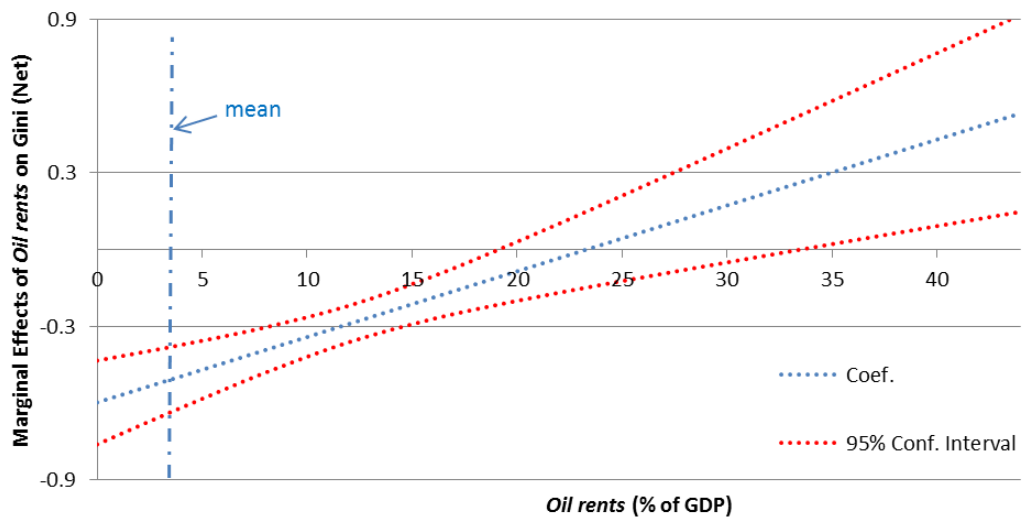

Note: Marginal effects based on specification 1 of Table 6. All other regressors at their mean values.

FIGURE 2. *Marginal Effects of Oil on Gini (Net) for Different Values of Oil Abundance (Specification 2, Table 6)*

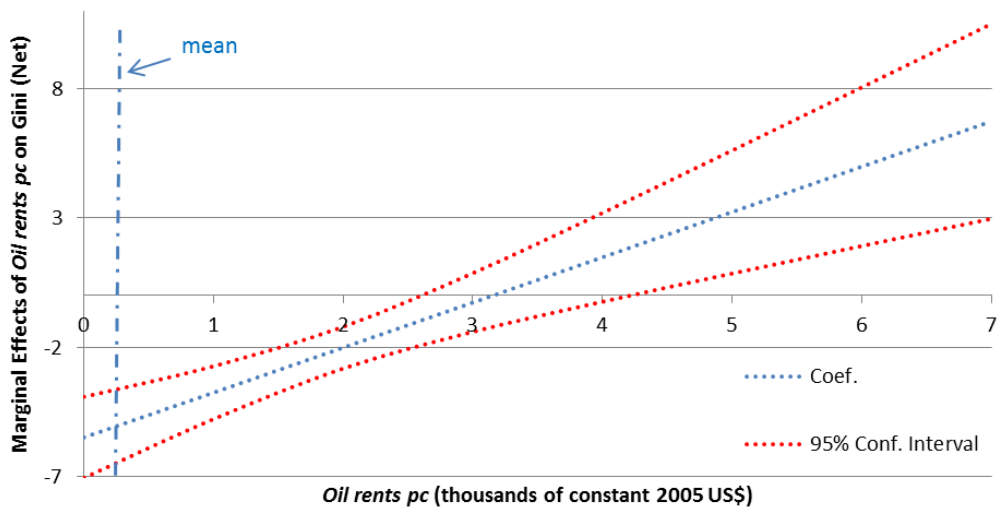

Note: Marginal effects based on specification 2 of Table 6. All other regressors at their mean values.

FIGURE 3. *Marginal Effects of Oil on Gini (Net) for Different Values of Oil Dependence (Latitude, Specification 1, Table 7)*

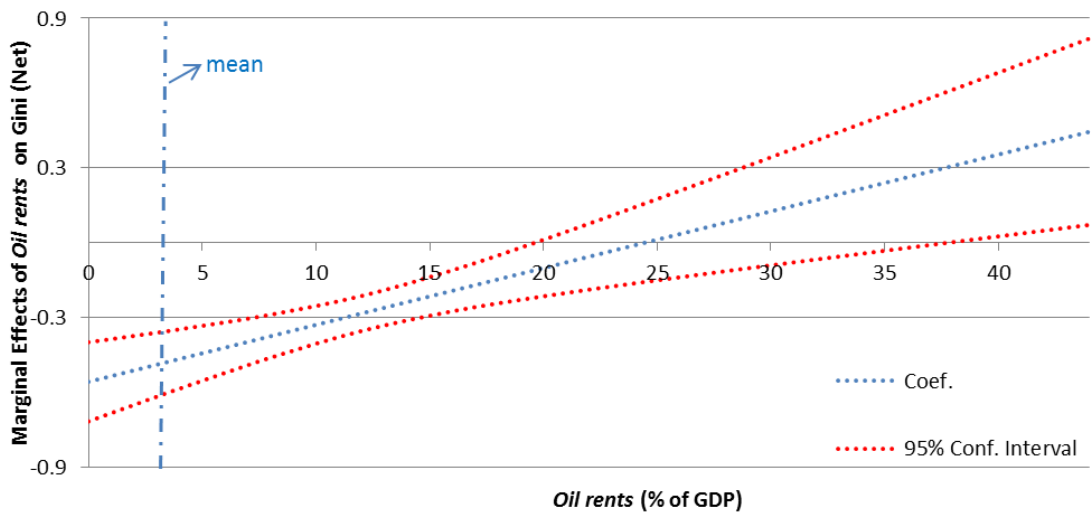

Note: Marginal effects based on specification 1 of Table 7. All other regressors at their mean values.

FIGURE 4. *Marginal Effects of Oil on Gini (Net) for Different Values of Oil Abundance (Latitude, Specification 2, Table 7)*

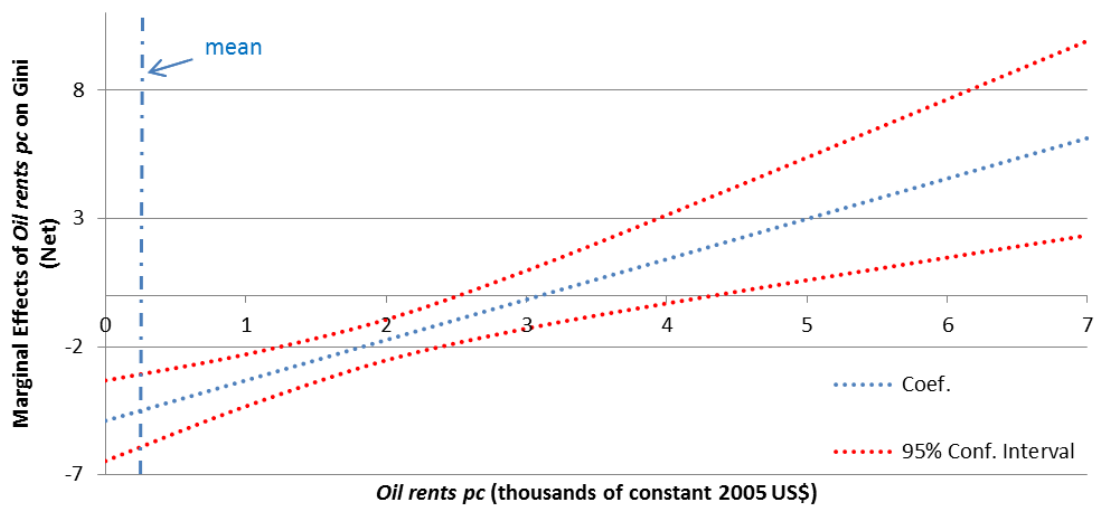

Note: Marginal effects based on specification 2 of Table 7. All other regressors at their mean values.

**Appendix 1: Pairwise Correlations for the Joint Sample of Reporting and Non-Reporting Countries**

| | Gini | Log Income | Latitude | Land locked | Oil rents | Oil rents pc | Oil reserves pc | Institutions | Agriculture | Fractionalisation | Openness | Rep. behav. | Data Rep. Index |
|---|---|---|---|---|---|---|---|---|---|---|---|---|---|
| *Gini* | 1 | | | | | | | | | | | | |
| *Log Income* | -0.49 | 1 | | | | | | | | | | | |
| *Latitude* | -0.62 | 0.65 | 1 | | | | | | | | | | |
| *Landlocked* | 0.11 | -0.31 | -0.09 | 1 | | | | | | | | | |
| *Oil rents* | 0.1 | 0.13 | -0.12 | -0.17 | 1 | | | | | | | | |
| *Oil rents pc* | -0.13 | 0.26 | 0.06 | -0.09 | 0.49 | 1 | | | | | | | |
| *Oil reserves pc* | -0.17 | 0.32 | 0.08 | -0.13 | 0.54 | 0.72 | 1 | | | | | | |
| *Institutions* | -0.33 | 0.44 | 0.38 | -0.12 | -0.29 | -0.13 | -0.27 | 1 | | | | | |
| *Agriculture* | -0.3 | -0.04 | 0.22 | 0.01 | -0.24 | -0.15 | -0.23 | 0.15 | 1 | | | | |
| *Fractionalisation* | 0.44 | -0.49 | -0.56 | 0.14 | 0.09 | -0.05 | -0.09 | -0.27 | -0.23 | 1 | | | |
| *Openness* | -0.03 | 0.22 | 0 | -0.02 | 0.12 | 0.12 | 0.16 | -0.01 | -0.21 | -0.16 | 1 | | |
| *Reporting behaviour* | -0.4 | 0.6 | 0.47 | -0.23 | -0.19 | -0.06 | -0.1 | 0.56 | 0.16 | -0.32 | -0.03 | 1 | |
| *Data reporting index* | 0.02 | 0.18 | 0.13 | -0.14 | -0.1 | -0.07 | -0.13 | 0.08 | 0.09 | -0.05 | -0.19 | 0.22 | 1 |

Note: A detailed description of all variables is provided in Table 1. Note: This table is based on the observations included in specification (6) of Table 2.

**Appendix 2: Pairwise Correlations for the Sample of Reporting Countries**

|  | Gini | Log Income | Latitude | Land locked | Oil rents | Oil rents pc | Oil reserves pc | Institutions | Agriculture | Fractionalisation | Openness | Rep. behav. | Data Rep. Index |
|---|---|---|---|---|---|---|---|---|---|---|---|---|---|
| *Gini* | 1 | | | | | | | | | | | | |
| *Log Income* | -0.67 | 1 | | | | | | | | | | | |
| *Latitude* | -0.75 | 0.65 | 1 | | | | | | | | | | |
| *Landlocked* | -0.07 | 0 | 0.12 | 1 | | | | | | | | | |
| *Oil rents* | 0.14 | -0.18 | -0.28 | -0.12 | 1 | | | | | | | | |
| *Oil rents pc* | -0.16 | 0.2 | 0.16 | -0.07 | 0.4 | 1 | | | | | | | |
| *Oil reserves pc* | -0.18 | 0.23 | 0.15 | -0.09 | 0.51 | 0.8 | 1 | | | | | | |
| *Institutions* | -0.38 | 0.58 | 0.42 | -0.02 | -0.15 | 0.11 | 0.13 | 1 | | | | | |
| *Agriculture* | -0.34 | -0.06 | 0.27 | 0.07 | -0.1 | -0.12 | -0.21 | 0.06 | 1 | | | | |
| *Fractionalisation* | 0.58 | -0.53 | -0.58 | 0.01 | 0.35 | -0.06 | -0.09 | -0.38 | -0.23 | 1 | | | |
| *Openness* | -0.14 | 0.18 | 0 | 0.17 | -0.02 | 0.06 | 0.09 | -0.07 | -0.1 | -0.2 | 1 | | |
| *Reporting behaviour* | n/a | n/a | n/a | n/a | n/a | n/a | n/a | n/a | n/a | n/a | n/a | n/a | |
| *Data reporting index* | 0.09 | -0.18 | -0.16 | -0.06 | 0 | -0.1 | -0.13 | -0.17 | -0.06 | 0.13 | -0.23 | n/a | 1 |

Note: A detailed description of all variables is provided in Table 1. Note: This table is based on the observations included in specification (6) of Table 2.

**Appendix 3: Observations and Mean Values per Country**

| Country | Freq. | % | Net Gini | Oil rents | Oil rents pc | Oil reserves pc |
|---|---|---|---|---|---|---|
| Argentina | 24 | 1.8 | 43 | 3.40 | 0.14 | 0.06 |
| Australia | 29 | 3.9 | 30 | 1.47 | 0.26 | 0.26 |
| Austria | 26 | 5.9 | 27 | 0.20 | 0.04 | 0.01 |
| Bangladesh | 19 | 7.3 | 37 | 0.79 | 0.00 | 0.00 |
| Belgium | 9 | 7.9 | 26 | 0.00 | 0.00 | 0.00 |
| Brazil | 24 | 9.7 | 51 | 0.83 | 0.02 | 0.05 |
| Canada | 29 | 11.9 | 29 | 2.28 | 0.43 | 0.38 |
| Chile | 24 | 13.6 | 50 | 0.59 | 0.02 | 0.01 |
| China | 24 | 15.4 | 41 | 4.24 | 0.02 | 0.02 |
| Colombia | 24 | 17.2 | 49 | 3.98 | 0.06 | 0.09 |
| Costa Rica | 24 | 19.0 | 42 | 0.00 | 0.00 | 0.00 |
| Denmark | 29 | 21.1 | 24 | 0.58 | 0.11 | 0.30 |
| Dominican Republic | 23 | 22.8 | 45 | 0.00 | 0.00 | 0.00 |
| Ecuador | 23 | 24.6 | 49 | 7.70 | 0.12 | 0.45 |
| Finland | 29 | 26.7 | 23 | 0.00 | 0.00 | 0.00 |
| France | 29 | 28.9 | 29 | 0.06 | 0.01 | 0.00 |
| Gambia | 12 | 29.7 | 53 | 0.00 | 0.00 | 0.00 |
| Germany | 20 | 31.2 | 27 | 0.08 | 0.01 | 0.00 |
| Greece | 28 | 33.3 | 33 | 0.10 | 0.01 | 0.00 |
| Guatemala | 22 | 34.9 | 51 | 0.31 | 0.00 | 0.02 |
| Honduras | 21 | 36.5 | 50 | 0.00 | 0.00 | 0.00 |
| Hungary | 29 | 38.6 | 27 | 1.61 | 0.05 | 0.03 |
| India | 24 | 40.4 | 49 | 1.47 | 0.01 | 0.01 |
| Indonesia | 24 | 42.2 | 46 | 7.51 | 0.06 | 0.03 |
| Ireland | 29 | 44.4 | 32 | 0.22 | 0.02 | 0.00 |
| Israel | 29 | 46.5 | 33 | 0.10 | 0.01 | 0.00 |
| Italy | 29 | 48.7 | 32 | 0.16 | 0.02 | 0.01 |
| Japan | 29 | 50.8 | 27 | 0.01 | 0.00 | 0.00 |
| Kenya | 22 | 52.4 | 49 | 0.00 | 0.00 | 0.00 |
| Korean Republic | 29 | 54.6 | 33 | 0.09 | 0.00 | 0.00 |

Note: This table is based on the observations included in specification (6) of Table 2.

**Appendix 3 (cont.): Observations and Mean Values per Country**

| Country | Freq. | % | Net Gini | Oil rents | Oil rents pc | Oil reserves pc |
|---|---|---|---|---|---|---|
| Malaysia | 23 | 56.3 | 47 | 7.00 | 0.16 | 0.29 |
| Mauritius | 24 | 58.1 | 18 | 0.00 | 0.00 | 0.00 |
| Mexico | 24 | 59.9 | 46 | 5.34 | 0.20 | 0.32 |
| Morocco | 23 | 61.6 | 37 | 0.04 | 0.00 | 0.00 |
| Netherlands | 29 | 63.7 | 25 | 1.57 | 0.29 | 0.01 |
| New Zealand | 29 | 65.9 | 31 | 0.30 | 0.03 | 0.08 |
| Nicaragua | 17 | 67.1 | 49 | 0.00 | 0.00 | 0.00 |
| Nigeria | 24 | 68.9 | 47 | 25.03 | 0.11 | 0.31 |
| Norway | 29 | 71.1 | 24 | 7.69 | 1.85 | 4.25 |
| Panama | 24 | 72.8 | 50 | 0.00 | 0.00 | 0.00 |
| Paraguay | 9 | 73.5 | 50 | 0.00 | 0.00 | 0.00 |
| Peru | 24 | 75.3 | 52 | 3.14 | 0.06 | 0.04 |
| Poland | 13 | 76.3 | 30 | 0.16 | 0.00 | 0.00 |
| Portugal | 29 | 78.4 | 31 | 0.00 | 0.00 | 0.00 |
| Singapore | 29 | 80.6 | 42 | 0.00 | 0.00 | 0.00 |
| South Africa | 24 | 82.3 | 55 | 0.90 | 0.03 | 0.00 |
| Spain | 29 | 84.5 | 32 | 0.08 | 0.01 | 0.00 |
| Sweden | 29 | 86.6 | 22 | 0.00 | 0.00 | 0.00 |
| Switzerland | 29 | 88.8 | 30 | 0.00 | 0.00 | 0.00 |
| Trinidad and Tobago | 21 | 90.4 | 37 | 19.26 | 1.28 | 0.92 |
| United Kingdom | 29 | 92.5 | 32 | 1.95 | 0.29 | 0.13 |
| United States | 29 | 94.7 | 35 | 1.35 | 0.34 | 0.10 |
| Uruguay | 24 | 96.4 | 43 | 0.00 | 0.00 | 0.00 |
| Venezuela | 24 | 98.2 | 41 | 17.84 | 0.76 | 1.96 |
| Zambia | 24 | 100.0 | 55 | 0.07 | 0.00 | 0.00 |

Note: This table is based on the observations included in specification (6) of Table 2.

## Appendix 4: Table 2 Re-estimated Using Fixed Effects

TABLE A1. *Effects of Oil Dependence on Net Income Inequality (Fixed Effects)*

| Dependent variable: | Gini (1) | Gini (2) | Gini (3) | Gini (4) | Gini (5) |
|---|---|---|---|---|---|
| *Log Income* | 5.193** (2.275) | 4.954** (2.064) | 5.543** (2.247) | 5.128* (2.945) | 5.092* (2.991) |
| *Oil rents* | -0.223** (0.087) | -0.468** (0.289) | -0.453** (0.181) | -0.910*** (0.205) | -0.896** (0.316) |
| *Oil rents (sq)* | | 0.005* (0.003) | 0.005* (0.03) | 0.017*** (0.006) | 0.016*** (0.006) |
| *Institutions* | | | 0.063 (0.133) | 0.059 (0.152) | 0.057 (0.147) |
| *Agriculture* | | | | -0.306** (0.122) | -0.301** (0.139) |
| *Trade Openness* | | | | | 0.923 (1.374) |
| $R^2$ adjusted | 0.203 | 0.198 | 0.213 | 0.258 | 0.260 |
| $N$ | 1935 | 1935 | 1592 | 1348 | 1348 |
| Countries | 81 | 81 | 75 | 55 | 55 |

Note: Robust standard errors of coefficients in parentheses. Superscripts *, **, *** correspond to a 10, 5 and 1% level of significance. Year dummies included in all specifications. A detailed description of all variables is provided in Table 1.

**Appendix 5: Conditional Marginal Effects for Heckman Selection Models**

The following is a general formula for the conditional marginal effect in the Heckman Selection Model (see Frondel and Vance, 2009), where $x_1$ is the variable for which the marginal effect is calculated. This formula is general in the sense that it can also be used to calculate the marginal effect even under the presence of an interaction term, such as $x_1 x_2$. When there is no interaction term, this formula reverts to the simple marginal effect because $\beta_{12} = \tau_{12} = 0$.

$$\frac{\partial E}{\partial x_1} = (\beta_1 + \beta_{12} x_2) + \beta_\lambda \cdot \delta(u_1) \cdot (\tau_1 + \tau_{12} x_2), \tag{A.1}$$

where

$u_1 = \hat{\tau}_1 Data\_reporting\_index_{it} + \hat{\tau}_2 Oil_{i(t-5)} + \hat{\boldsymbol{\tau}}_3' \boldsymbol{Z}_{it}$, and

$$\delta(u_1) = \lambda'(u_1) = \frac{-u_1 \phi(u_1) \Phi(u_1) - \phi^2(u_1)}{\Phi^2(u_1)} = -[\lambda(u_1)]^2 - u_1 \cdot \lambda(u_1).$$